# POCFormer: A LIGHTWEIGHT TRANSFOMER ARCHITECTURE FOR DETECTION OF COVID-19 USING POINT OF CARE ULTRASOUND

Shehan Perera[1], Srikar Adhikari[2], Alper Yilmaz[1]

[1]Photogrammetric Computer Vision Lab. The Ohio State University, Columbus, Ohio
[2]Emergency Medicine, The University of Arizona, Tucson, Arizona

## ABSTRACT

The rapid and seemingly endless expansion of COVID-19 can be traced back to the inefficiency and shortage of testing kits that offer accurate results in a timely manner. An emerging popular technique, which adopts improvements made in mobile ultrasound technology, allows for healthcare professionals to conduct rapid screenings on a large scale. We present an image-based solution that aims at automating the testing process which allows for rapid mass testing to be conducted with or without a trained medical professional that can be applied to rural environment and third world countries. Our contributions towards rapid large-scale testing includes a novel deep learning architecture capable of analyzing ultrasound data that can run in real time and significantly improve the current state-of-the-art detection accuracies using image based COVID-19 detection.

## 1. INTRODUCTION

Coronavirus disease 2019 (COVID-19) is a relatively new virus that has caused a recent outbreak of respiratory illnesses starting from an isolated event to a global pandemic. As of January 13 2021, there are over 22.8 million confirmed COVID-19 cases in the United States and over 92 million worldwide. In the United States alone, over 381,000 Americans have died from COVID-19, with no end in sight (https://coronavirus.jhu.edu/map.html).

Aside from multiple nonapproved alternatives, there are currently, two types of tests that are conducted by healthcare professionals– Molecular tests and Serology tests. The molecular tests help diagnose an active coronavirus infection in a patient. The ideal diagnostic test and the "gold standard" according to the United States Center for Disease Control (CDC) is the Reverse Transcription Polymerase Chain Reaction (RT-PCR) [1,2]. Although highly accurate, methods such as RT-PCR do not meet the speed requirements needed for testing on a large scale.

Recently Point-of-Care (POC) ultrasound devices have started to be adopted by healthcare professionals due to their reliability and portability. An emerging popular technique, which adopts improvements made in mobile ultrasound technology, allows for healthcare professionals to conduct rapid screenings on a large scale. As it stands, the screening of a new patient, with the use of a mobile ultrasound device takes about 13 minutes, with the caveat that it requires a highly trained professional to interpret the results generated by the device. However, the experience and expertise to interpret ultrasound findings are not universally available. With the combination of deep learning and computer vision, we introduce a novel approach that significantly improve testing time by automating the detection of COVID-19 with an accuracy comparable to a trained sonographer. The approach differentiates COVID-19 against bacterial pneumonia for better diagnosis. Our contributions include:

• A novel network architecture to diagnose COVID-19 that relies on ultrasound clips.

• An approach that improves on recent advancements in Point of Care ultrasound technology.

• A solution that greatly improves diagnosis of COVID-19 using single ultrasound frames.

• Greatly reduces testing time with real-time testing capabilities.

## 2. RELATED RESEARCH

Since the outbreak of COVID-19 researchers have proposed a number of methods used to diagnose patients. Without exception, the most common imaging modalities used are the chest x-rays (CXR), and lung CT scans [3]. Chest x-rays are generally used to evaluate respiratory symptoms since they are capable of diagnosing many conditions at different stages of a lung disease. For detection of COVID-19, CXR is not reliable, particularly in early stage.

CT scans, although highly sensitive in detecting COVID-19 [4, 5], are less efficient from a data collection point of view due to the required sanitization of the equipment and the room after each patient, which usually requires 60 to 120 minutes. In addition, many facilities globally are unable to provide CT scans as a viable primary diagnostic tool either because it is expensive or that the facility does not have a CT Scanner. With the speed and accuracy being the major bottleneck in rapid mass testing, there is a need for fast accurate tool to diagnose and monitor development of COVID-19.

Point of care lung ultrasound is readily available in most acute care settings, can be rapidly performed bedside with no exposure to ionizing radiation and does not need the area, staff and time required for CT imaging. POC lung ultrasound has been shown to be superior to CXR in its ability to detect pulmonary abnormalities in patients with both COVID-19 and non-coronavirus pulmonary infections

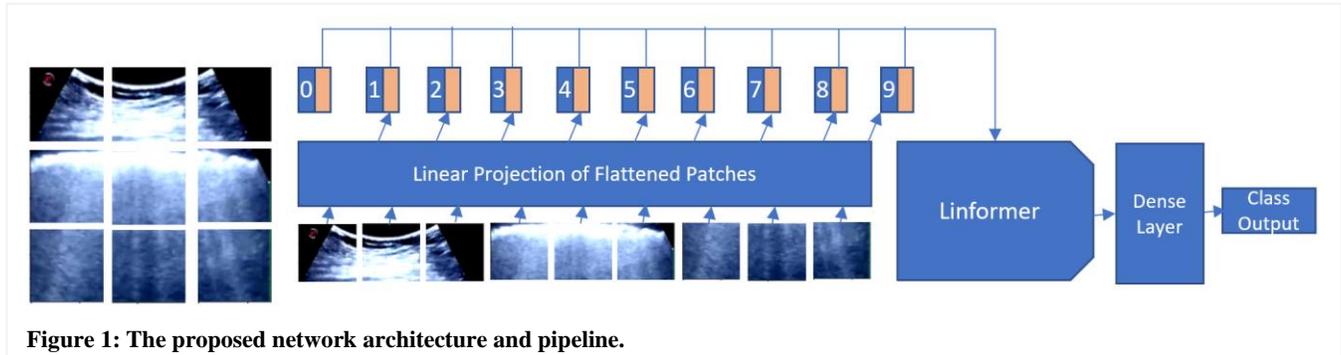

**Figure 1:** The proposed network architecture and pipeline.

[6]. It offers a unique strategy for screening and risk stratification of suspected COVID-19 patient without consuming utilizing significant healthcare resources. After the initial evaluation, serial ultrasound examinations may be helpful to track the clinical trajectory of COVID-19, direct appropriate therapy and assess the clinical response to the therapeutic interventions. In the clinical trajectory of COVID-19, direct appropriate therapy and assess the clinical response to the therapeutic interventions. In addition, portable ultrasound devices are easier to decontaminate than portable CXR or CT equipment. Recently, the American College of Radiology recommended that CT should not be used to screen for or diagnose COVID-19 and that it should only be used selectively in hospitalized, symptomatic patients given the risk of infection transmission posed to hospital personnel and patients. Ultrasound similar to CT scan is capable of providing information that can be used for accurate diagnosis of COVID-19[6]. In a recent study, comparing CT scans to ultrasounds for monitoring COVID-19 patients, [7] suggested that the overlaps between the B-lines in the ultrasound versus the ground glass opacities (GGO) visible in CT scans proved the use case of ultrasounds in evaluating patients over the CT scans. Just as accurate as CT scans but order of magnitudes faster compared to x-ray and CT many physicians are adopting ultrasound globally as a viable way to provide mass rapid testing capability to people in need. POC lung ultrasound can help healthcare providers distinguish COVID-19 from other causes of dyspnea. Although a comprehensive dataset of ultrasound for COVID-19 does not exist, few researchers have published work where deep learning has been used to automate the detection of COVID-19. POCOVID-Net by [8] proposed a frame-by-frame classification of lung ultrasound images and boasted a balanced accuracy of 82% which was a major improvement over COVID-Net [9] which analyzed chest CT scans to detect the presence of COVID-19 and reached an overall balanced accuracy of 63%. Like POCOVID-Net, optimized variations of VGG and NasNet mobile have also been proposed [10] for COVID-19 detection which achieved an accuracy 89% and 79% respectively.

## 3. METHODOLOGY

The proposed lightweight transformer network architecture shown in Figure 1 is composed of a vision transformer which aids in pre-processing the input image and a linear transformer as the primary feature encoder. The overall detection pipeline is designed to be simple, effective, and efficient with around 2 million parameters.

### 3.1. DATA AND IMAGE PREPROCESSING

The data used for the training, validation and testing were obtained by the only open-sourced lung ultrasound for COVID-19 dataset [8]. The dataset consists of lung ultrasound images and video that were donated by medical institutions around the world. The training and testing dataset were created by extracting frames from lung ultrasound video data and each of the three classes Covid-19, Pneumonia and Healthy patient classes were balanced during the training, validation and testing dataset creation by introducing a weighting mechanism into the loss function to enforce emphasis on the classes that are underrepresented.

$$\text{weight\_vector} = \frac{1}{X/min(X)},$$

where **X** contains the number of training samples pre class.

Ultrasound image frames contain features, such as A-lines and B-lines, that are manifested in the higher frequency content. To standardize the input images collected from various sources each frame is resized to a standard 224x224 image. Each original 8bit image is normalized to be in the rage between 0 and 1 prior to adjusting the mean and standard deviation of each image. The mean, $\mu$, and standard deviation, $\sigma$, are calculated as:

$$\mu = (\frac{1}{n}\sum_{i=i}^{n} x_i), \qquad \sigma = \sqrt{\frac{1}{N}\sum_{i=1}^{N}(x_i - \mu)^2}$$

where $x_i = [R_i \; G_i \; B_i]^\top$ denoting the red, green, and blue channels.

### 3.2. TRANSFORMER NETWORK

In addition to improving classification accuracies compared to typical standard feature extractors dominated by CNN's in medical imaging, we proposed the use of a transformer-based image classification approach. The Vision Transformer [11] shown in Figure 1, feature extractor is capable of rapidly extracting accurate features that distinguish covid-19, pneumonia and a healthy patient within a few epochs. The rapid adaptability of the transformer architecture a can be seen during the training and evaluation process since most of the data in the dataset are from various sources that include different ultrasound machines making each the dataset highly diverse.

To ensure the proposed transformer architecture is memory and time efficient the standard transformer model [12,13] is replaced by a linear transformer model (Linformer)[14] shown in Figure 2. The replacement of the Linformer architecture allows the vision transformer to perform the internal self-attention mechanism that originally has space time complexity of $O(n^2)$ by a mechanism with smaller complexity of $O(n)$. The Linformer's reduced complexity is achieved by techniques including but not limited to approximating the original self-attention mechanism by a low-rank matrix, parameter sharing between each head, and incorporating two linear projections during computation of the key value pairs. These mechanisms make the final network to remain compact while delivering feature extraction and self-attention capabilities of a much larger 345 million parameter transformer such as Bidirectional Encoder Representations from Transformers [13].

The vision transformer (ViT) model seen in Figure 1 was chosen as the primary feature extractor in the pipeline in combination with the Linformer model to combine the benefits of attention and efficiency in order to develop a system that can be used in a real time mobile setting. The ViT architecture allows for minimal alternation of the standard transformer used in natural language processing by combining a sequence of image patches of the input image with an embedding layer that combines the positional information with the input sequences. The results show that proposed architecture is highly adaptable and efficient that it is capable of learning from few training data points is invaluable in situations where data is scare.

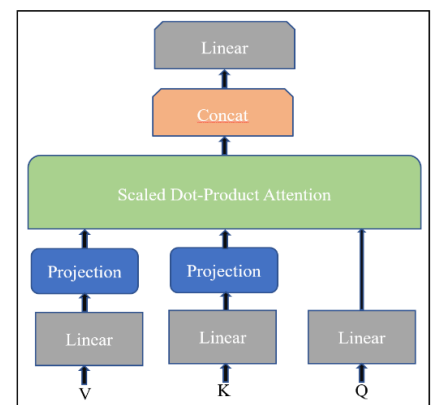

**Figure 2: A single head of the Linformer architecture.**

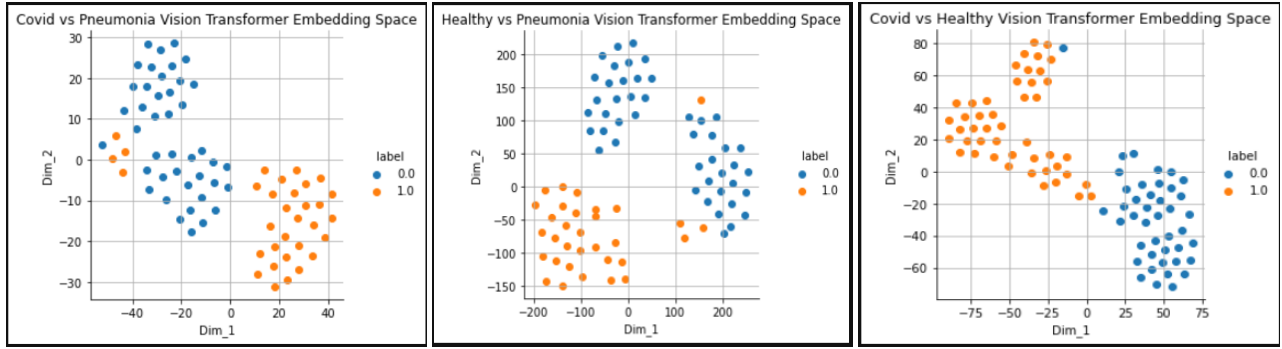

**Figure 3:** TSNE visualization of the binary classification embedding space. The plots demonstrate the linear separability across classes in low dimensional space.

The cost function during training uses the cross-entropy loss to compute the probability of error between the predicted output class and the target class. The cross-entropy loss function can be expressed as:

$$\text{Cross Entropy Loss} = -\sum_{c=1}^{M} y_{o,c} \log(p_{o,c}),$$

where, p - predicted probability observation o is of class c and y is a binary indicator (0 or 1) if class label $c$ is the correct classification for observation $o$.

### 3.3. HYPERPARAMETER SEARCH

Significant effort on hyperparameter search was conducted. Since the ViT network in conjunction with the Linformer model performs the final classification of the input image the hyperparameter search was needed to be done simultaneously so that each component compliments each other. The final hyperparameters for the binary classification and multiclass classification are shown in Table 1 below.

### 4. EXPERIMENTS

The model training, validation and testing pipelines were implemented using the PyTorch framework. All experiments presented are designed, developed, and tested on a workstation running an, Intel Core i7-9700K CPU at 4.7GHz, and an RTX 2080 GPU with 8GB RAM. To demonstrate the networks capability to learn from few data points, the networks were trained with a dataset of size ~400 images per class which is a much smaller subset of the full dataset. This process not only helped demonstrate the learning efficiency of the network, it also ensured the training dataset is well balanced. The training dataset is broken down into 3 categories 1) confirmed COVID-19, 2) confirmed bacterial pneumonia and 3) confirmed healthy. To evaluate the performance of the proposed model and understand its diagnosis capability, we conducted two sets of experiments, binary and multiclass classification.

### 4.1 BINARY

Binary classification consisted of three separate experiments with no change to the network architecture or the model development process. The purpose of binary classification is threefold. Mainly the binary classification experiments help to demonstrate the linear separability of the input lung ultrasounds in low dimensional space as seen in Figure 3. Linear separability in low dimensional space demonstrates the networks ability to separate each class and perform accurate classification. Secondly, by observing the network architectures embedding space, a deeper understanding of the effects of hyperparameters can be obtained which helps to improve the network architecture for maximum performance. Finally, the binary classification experiments were designed to ensure that no single category dominates the results produced by the network. The experiments conducted are presented below.

1) COVID-19 versus healthy – Tests algorithm's ability to distinguish between a healthy and COVID-19 patient.
2) COVID-19 versus bacterial pneumonia – Tests algorithm's ability to distinguish between a COVID-19 and bacterial pneumonia
3) Bacterial pneumonia versus healthy – Tests algorithm's ability to distinguish between a healthy and bacterial pneumonia patient

### 4.2 MULTICLASS

To evaluate the networks performance in a real-world scenario a final multiclass classification experiment was conducted. Multiclass classification tests the classification performance between COVID-19, Bacterial pneumonia, and a healthy patient. With a minor modification of the binary classification network architecture, we can experimentally demonstrate the network architecture can be adapted to improve the state-of-the-art in lung ultrasound classification, while ensuring real time capability.

### 5. RESULTS

The binary diagnosis results are tabulated in Table 2 and multiclass diagnosis results are tabulated in Table 3. We show that the proposed classification network architecture with a 2.8 million parameters count for binary classification and 6.9 million parameters count for multiclass classification performs as well or better compared to approaches with networks with parameters ranging from less than 10 million to greater than 130 million parameters.

In the case of binary classification between, which was performed as an experiment to showcase the proposed network's ability to linearly separate the input lung ultrasound frames in a low dimensional space Figure 3. The binary classification task which on its own demonstrates the efficiency of the proposed architecture it also served as a foundation to explain the performance seen in the multiclass classification scenario and more importantly it allowed us to identify the effects of certain hyperparameters which significantly aided in the development of the final network architecture.

| Model | Layers | Hidden Size | MLP size | Heads | Patch Size | Param Count |
|---|---|---|---|---|---|---|
| Binary Model | 12 | 64 | 128 | 8 | 32 | 2.8M |
| Multiclass Model | 32 | 64 | 128 | 8 | 32 | 6.9M |

**Table 1:** Hyperparameters of the binary and multiclass model.

|  | Pr. | Ac. | F1 | Sen. | Sp. |
|---|---|---|---|---|---|
| COVID-19 vs Healthy | 0.87 | 0.91 | 0.92 | 0.83 | 0.97 |
| COVID-19 vs Pneumonia | 0.93 | 0.95 | 0.93 | 1.00 | 0.88 |
| Pneumonia vs Healthy | 0.94 | 0.95 | 0.93 | 1.00 | 0.80 |

**Table 2:** The binary classification performance metrics. Pr, Ac, Sen and Sp respectively denote precision, accuracy, sensitivity and specificity.

Table 3 demonstrate the classification improvements made by the proposed network architecture over POCOVID-Net and optimized versions of VGG and NasNet architectures. The multiclass classification accuracies of the proposed network demonstrate a significant improvement in precision, recall, f1 sensitivity, specificity, and accuracy metrics against state-of-the-art network architectures. The bold face results in the shows the best result for the category.

## 5. DISCUSSION

The proposed network architecture achieves state of the art results in ultrasound based COVID-19 detection. The binary classification variation of the proposed network with roughly 2 million parameters, which is about half of MobileNetv2[1] demonstrates an average accuracy of 91% at 70 frames per second. The size and performance of the network allows it to be deployed on edge or devices with low computational power, such as smart phones. Similarly, the multiclass variation of the proposed architecture with roughly 6.9 million parameters performs at or above the classification performances of much larger networks with 38.4 frames per second. The performance improvements in classification and efficiency can be attributed to improvements introduced by the visual transformer in combination with the Linformer feature extractor. This combination demonstrates that high efficiency and high performance can be achieved with compact networks.

## 6. CONCLUSIONS

We proposed a new architecture composed of vision and linear transformers that demonstrates the efficiency and performance improvements in lung ultrasound classification over state of the art. With POC ultrasound devices readily available in the market, our goal is to show possibility of rapid mass testing is possible with or without a trained medical professional. The experiments on a well-known open sourced dataset show that the parameter set can be successfully learned to achieve results over 90% accuracy and precision.

| Architecture | Class | Recall | Precision | F1-Score | Specificity | Sensitivity |
|---|---|---|---|---|---|---|
| **POCOVIDNet [8]** Acc: 0.89 Param: +130M | COVID-19 | n/a | 0.88 | 0.92 | 0.79 | **0.96** |
|  | Pneumonia | n/a | 0.95 | 0.94 | 0.98 | 0.93 |
|  | Healthy | n/a | 0.78 | 0.62 | **0.98** | 0.55 |
| **Pruned VGG [10]** Acc: 0.90 Param: 14.7M | COVID-19 | 0.89 | 0.91 | 0.90 | 0.92 | n/a |
|  | Pneumonia | 0.94 | **0.93** | 0.94 | 0.97 | n/a |
|  | Healthy | 0.85 | 0.83 | 0.83 | 0.95 | n/a |
| **NasNet Mobile [10]** Acc: 0.76 Param: 4.8M | COVID-19 | 0.87 | 0.74 | 0.80 | 0.72 | n/a |
|  | Pneumonia | 0.79 | 0.88 | 0.83 | 0.96 | n/a |
|  | Healthy | 0.47 | 0.61 | 0.53 | 0.92 | n/a |
| **Ours** Acc: 0.939 Param: 6.9M | COVID-19 | **0.96** | 0.90 | **0.93** | **0.98** | 0.90 |
|  | Pneumonia | **0.94** | **0.99** | **0.96** | 0.96 | **0.98** |
|  | Healthy | **0.92** | **0.93** | **0.93** | 0.96 | **0.92** |

**Table 2:** Multiclass classification performance compared to other network architectures developed on the same open-sourced lung ultrasound dataset